# Performance Limits Projection of Black Phosphorous Field-Effect Transistors

Kai-Tak Lam, *Member, IEEE*, Zhipeng Dong and Jing Guo, *Senior Member, IEEE*

*Abstract*—Ballistic device performance of monolayer black phosphorous (BP) field-effect transistors (FET) is investigated in this work. Due to the anisotropic effect mass of the carriers, the ON-state current is dependent on the transport direction. The effective masses are lower in the "armchair" direction which provides higher drive current at the same biasing. The degree of anisotropy is higher for the holes, which improves the performance of p-type devices. The intrinsic delay of 20 nm BP FETs is in the range of 50 fs at ON-/OFF-current ratio of 4 orders. Monolayer BP FETs outperform both $MoS_2$ and Si FETs for both n- and p-type devices in terms of ballistic performance limits, due to highly anisotropic band structure.

## I. Introduction

THE research in ultra-thin body electron devices have seen the introduction of atomic scale two-dimensional (2D) materials, from the few atomic layer III-V semiconductors [1] to single atomic layer graphene [2]. Other 2D materials such as hexagonal boron nitride and various transition metal dichalcogenides have been investigated [3-5], which, unlike graphene, have a substantial band gap and are preferable for electron device applications, but suffer with relatively low mobility values. Another layered material known as black phosphorus (BP) was recently demonstrated as FET devices [6-8] with an ON-/OFF-current ratio of 5 orders and a field-effect mobility of ~1,000 $cm^2/Vs$. Monolayer BP has a puckered honeycomb structure with a direct band gap of 1~2 eV and the effective masses have a high degree of anisotropy that is not observed in other 2D materials. This suggests that the device performance of BP FETs is highly dependent on the transport direction and hence, in this work, we investigate the influence of effective mass anisotropy on the device performance of BP FETs. We obtained the 2D conduction and valence bands within the first Brillouin zone of monolayer BP using density functional theory and simulated the device performance with an analytical ballistic MOSFET quantum transport model [9] coupled with a self-consistently solved Poisson potential based on the capacitance model. Due to the higher degree of anisotropy in the hole effective mass, the ON-state current ($I_{ON}$) of p-type device is higher than that of n-type device. The average carrier velocity and the intrinsic delay of the devices are extracted and the n- and p-type devices are found to have compatible characteristics. We have also simulated the device performance of MoS2 FETs under the same biasing conditions and found that BP FETs can outperform both n- and p-type devices.

## II. Methodology

The atomic structure of monolayer BP was optimized using density function theory (DFT) implemented in VASP [10]. The optB88-vdW functional [11] is used which results in a lattice constant of $a$ = 4.57 Å and $b$ = 3.33 Å. A choice of $c$ = 15 Å was made to ensure ample separations between BP layers during calculations. Four phosphorus atoms form a unit cell and their positions are listed in Table I. A top view of the atomic structure is shown in Fig. 1(a) and the corresponding band structure calculated using the HSE06 functional [12] is shown in Fig. 1(b). A 10×10×1 Monkhaust Pack k-point grid is used for DFT calculation and the effective masses for the electron and holes in different directions are extracted and summarized in Table I. A direct band gap of 1.5 eV was obtained which matches with previous results [13], and is preferred for a low OFF-state current ($I_{OFF}$) in transistor operation. For carrier transport simulation, dense electron dispersions are obtained from Wannier interpolation [14], shown as the line plots in Fig. 1(b) and as color plots in Fig. 1(c) and 1(d) for the conduction and valence bands respectively. An iso-energy contour line 0.05 eV from the band edges visualized the degree of effective mass anisotropy,

Manuscript received XXX XX, 2014. This work was supported by the NSF and ONR.
The authors are with the Department of Electrical and Computer Engineering, University of Florida, Gainesville, FL 32611, USA. Corresponding author: J. Guo (Tel: 1-352-392-0940; Email: guoj@ufl.edu)
.

which is higher for holes than electrons.

A double-gated planar device structure is simulated in this work, shown in the inset of Fig. 2(c). The gate oxide is 3 nm $ZrO_2$ ($\varepsilon_r$ = 25, $C_{OX}$ = 73.8 fF/μm$^2$) and the gate control is assumed to be perfect with short channel effect ignored. The equilibrium and non-equilibrium charge densities $N_0 = \int_{-\infty}^{\infty} D(E)f(E-E_F)dE$, $N = \frac{1}{2}\int_{-\infty}^{\infty} D(E-U_{SC})[f(E-\mu_S) + f(E-\mu_D)]dE$, where $D(E)$ is the density of states (DOS) obtained via the calculated band structure and $f$ is the Fermi-Dirac distribution function with $E_F$ being the Fermi level of the system and $\mu_{S/D}$ being the chemical potentials at the source and drain terminals ($\mu_S = E_F$ and $\mu_D = E_F - V_D$, $V_D$ being the drain bias). The self-consistent potential $U_{SC}$ is obtained using the capacitance model, $U_{SC} = q^2(N-N_0)/2C_{ox} - qV_G$. The current $I_{DS}$ is evaluated from the difference between the fluxes from the source and drain terminals, related to the carrier group velocity $dE_{\vec{k}}/d\vec{k}$ at different transport directions. Average carrier velocity, $\langle v \rangle = I_{DS}/qN$ where $q$ is the elementary charge. The magnitude of the drain bias $|V_D|$ = 0.6 V and the $E_F$ is adjusted such that the $I_{OFF}$ at $V_G$ = 0 V is at 100 nA/μm, with the $I_{ON}$ defined at $V_G = V_D$.

### III. RESULTS AND DISCUSSION

The device performance of monolayer BP FETs are summarized in Fig. 2. The angle resolved current characteristics depending on the transport direction are shown in Fig. 2(a) and 2(b) for the n-type and p-type devices respectively. Due to the much lower effective masses in the x-direction ["armchair" direction as shown in the top and side views in Fig. 1(a) and inset of Fig. 2(c)], the $I_{ON}$ decreases as the transport direction changes from x to y. We further extracted the current characteristics at the two orthogonal directions and plotted in Fig. 2(c) for easy comparison. Due to the higher anisotropy of the hole effective mass, the $I_{ON}$ is higher for the p-type device in the x-direction. While the effective mass in the transport directions are similar (0.17 and 0.16 for electrons and holes respectively), the larger hole DOS results in a larger current for p-type device. On the other hand, carrier transport in the y-direction suffers as the higher effective masses translate to a lower carrier injection velocity. The $I_{ON}$ for n-type and p-type devices reduce by 50% and 75% respectively when the transport direction changes from x to y.

The average carrier velocity and the intrinsic delay are also extracted for the different transport directions for the n- and p-type devices, summarized in Fig. 3. The average carrier velocity is related to the effective mass in the transport direction and the much heavier hole in the y-direction results in a lower average carrier velocity [cf. Fig. 3(a)]. The inset of Fig. 3(a) shows the decrease in average carrier velocity at ON-state as the transport direction changes from x to y-direction for both n- and p-type devices and it is observed that both devices have comparable average carrier velocity up to 45°, providing a certain degree of freedom for high performance devices. The intrinsic delay for a 20 nm device is extracted and shown in Fig. 3(b) where at an $I_{ON}/I_{OFF}$ ratio of 4 orders, both n- and p-type monolayer BP FETs delivers a ~50 fs intrinsic delay in the "armchair" transport direction. These device performance comparisons highlight the need to properly identify the orientation of the monolayer BP and apply the contacts in the correct direction for high performance devices.

Lastly, we carried out similar simulations using $MoS_2$ monolayer and ultra thin body (UTB) Si summarized in Fig. 4(a) and 4(b), assessed by using the same model and device structure as in [15]. For UTB Si, the thickness was assumed to be 3 nm with a confinement direction of (100) and a native oxide thickness of 0.4 nm. The nearly isotropic effective masses of monolayer $MoS_2$ result in similar current characteristics irrespective of transport directions. At a common $I_{OFF}$, the ballistic $I_{ON}$ of BP FETs in the "armchair" transport direction outperforms that of monolayer $MoS_2$ FETs by a factor of 1.57 and 1.89 for n- and p-type devices and Si FETs by 1.69 and 2.41.

The performance advantage can be attributed to highly anisotropic band structure of monolayer BP. The $I_{ON}$ of a transistor is determined by the product of the charge and average carrier velocity. New channel materials such as III-V semiconductors have small effective mass that can offer higher carrier injection velocity than Si but suffer from lower DOS. For materials with isotropic band structure, the tradeoff between carrier velocity and DOS limits the $I_{ON}$ and ballistic performance of FETs [16]. On the other hand, an anisotropic band structure, which has a small transport mass leading to a larger carrier injection velocity, and a large transverse mass leading to higher DOS, is expected to have superior device performance. Monolayer BP, whose electron masses differ by a factor of 10 and hole masses differ by a factor of 40 in orthogonal directions, is preferred in this regard. Finally, we simulated the change in $I_{ON}$ for few-layer BP FETs using effective mass values in [17], shown in Fig. 4(c) and 4(d). The $I_{ON}$ shows minimal changes as the layer increases from 1 to 5, except for the p-type device with y-direction transport. The hole effective mass in the y-direction reduces from 6.35 to 0.89 as the layer increases while that of electrons increases slightly from 1.12 to 1.18. The bandgap, however, decreases from ~1.5eV to ~0.59eV as the layer increases from 1 to 5, which can lead to considerable increase of the minimal leakage current.

## IV. Conclusion

We have investigated the ballistic device performance of monolayer black phosphorus (BP) field-effect transistors (FETs). Due to the highly anisotropic effective masses, the current characteristics are transport direction dependent. In the lower effective mass directions, the higher degree of anisotropy of the hole effect mass results in a higher ON-state current ($I_{ON}$) for the p-type device as a result of a higher density of states. Compared with similar monolayer MoS$_2$ and UTB Si devices, BP FETs achieved a higher $I_{ON}$ in the "armchair" transport direction due to the higher injection velocity and higher density of states. In the orthogonal direction, BP FETs gives comparable results. Our investigation highlights the potential of monolayer BP FETs as high performance devices and the importance of identifying the orientation of the material for optimal performances.

## Acknowledgement

The authors would like to thank Prof. F. Xia of Yale University and Dr. H. Wang of IBM for technical discussions.

TABLE I
MATERIAL PARAMETERS OF MONOLAYER BLACK PHOSPHOROUS

| Lattice (Å) | $a = 4.5694$ | $b = 3.3255$ | $c = 15$ |
|---|---|---|---|
| Atom positions (Å) | ($0.0860\,a$, $b/2$, $2.1360$) | ($0.4140\,a$, $0$, $2.1360$) | |
| | ($0.5860\,a$, $0$, $0$) | ($0.9140\,a$, $b/2$, $0$) | |
| Electron $m^*$ ($m_0$) | 0.17 (x) | 1.20 (y) | |
| Hole $m^*$ ($m_0$) | 0.16 (x) | 6.49 (y) | |

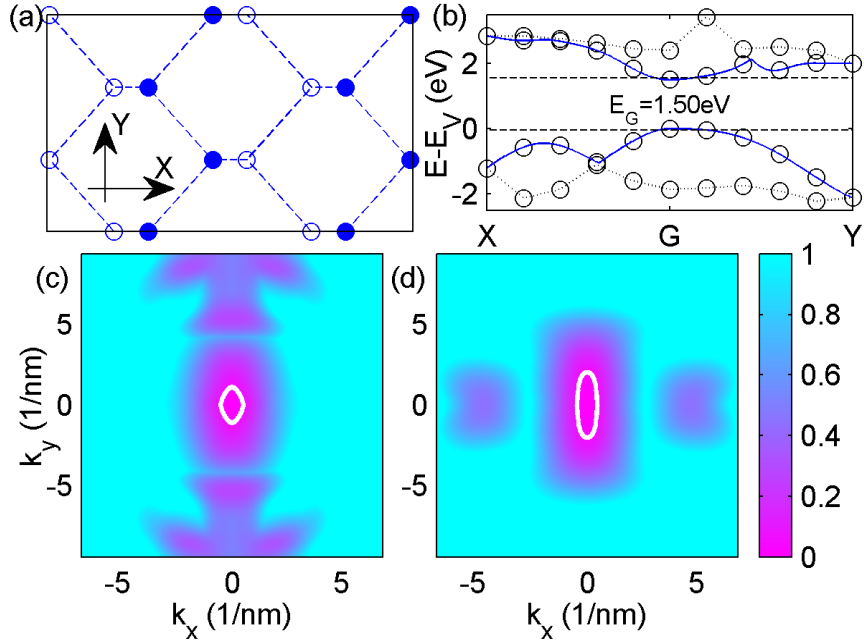

Fig. 1. (a) Top view of the atomic structure of monolayer black phosphorus (BP). The empty and filled dots denote different z-positions. (b) The band structure of monolayer BP plotted along the high-symmetry paths, calculated from the 10×10×1 Monkhaust Pack grid (markers) and from Wannier interpolation (lines), normalized to the valence band edge ($E_V$). The solid lines are the conduction and valence bands, whose two dimension representations are shown in (c) and (d) respectively. The color bar on the right applies to both (c) and (d) in |eV| and 0 represents the band edges.

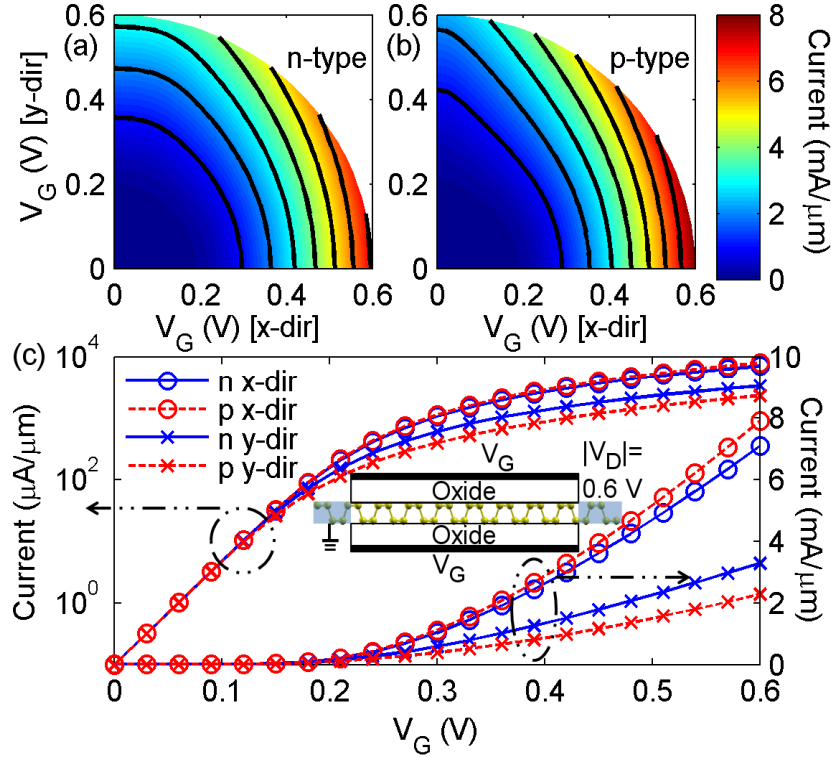

Fig. 2. The angle resolved current characteristics of monolayer BP FETs with $|V_D| = 0.6$ V for (a) n-type and (b) p-type devices, at different transport directions from 0° (x-dir) to 90° (y-dir). The color bar applies to both figures and the contour lines go from 1 to 8 mA/μm. (c) The current characteristics for the n-type (solid) and p-type (dash) devices at the x- (circle) and y-direction (cross) in log and linear scale. The inset shows the side view of the double-gated device structure in the x transport direction.

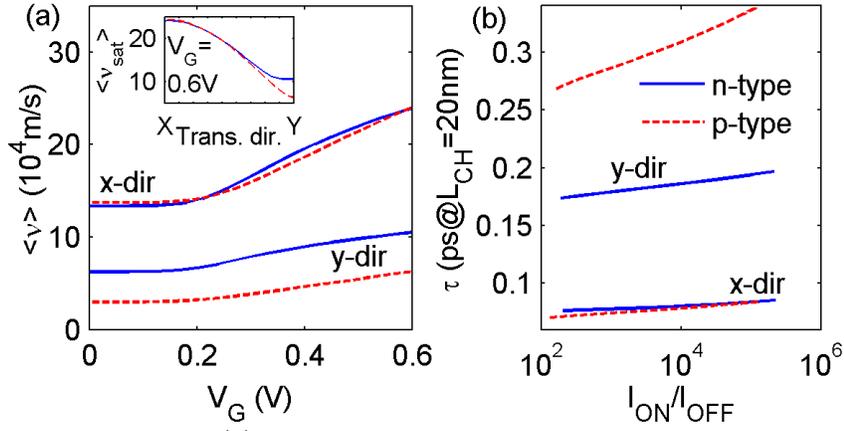

Fig. 3. (a) The average carrier velocity $\langle v \rangle$ for n- and p-type monolayer BP FETs in the x- and y-direction. The inset shows the change in average saturation velocity. (b) The intrinsic delay $\tau$ as a function of ON-/OFF-current ratio in x and y transport directions, with a channel length of 20 nm.

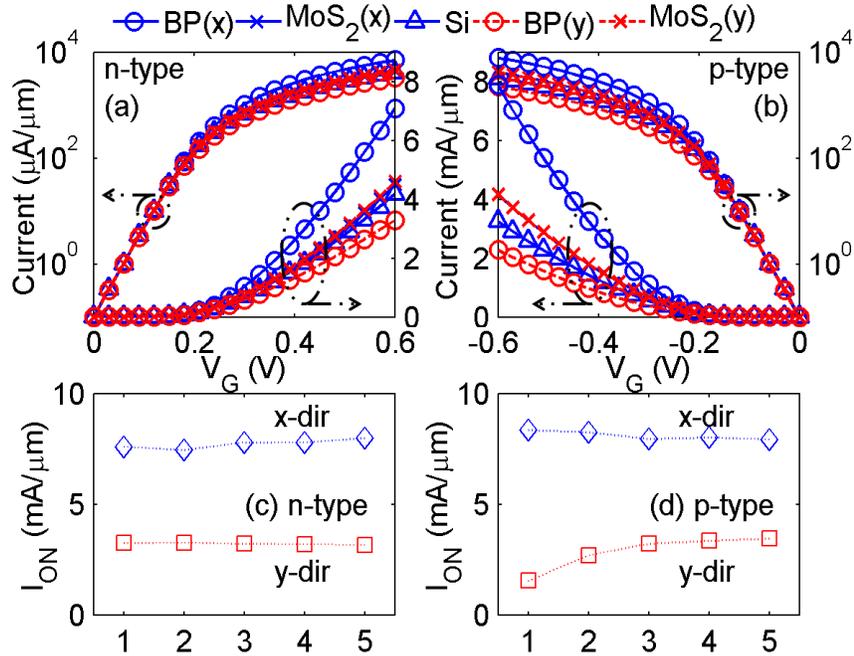

Fig. 4. Comparison between monolayer BP (circle), monolayer MoS$_2$ (cross) and ultra-thin body Si (triangle) for (a) n-type and (b) p-type devices in x- (solid) and y-direction (dash). The markers of MoS$_2$ devices are overlapped on each other, an indication of the isotropic effective masses. There is only one transport direction due to (100) confinement for the UTB Si devices. The ON-state current as a function of layer number in BP FETs for (c) n-type and (d) p-type devices for both x- and y-directions. The x-axis shows the layer number in both figures.